# Probing nanoscale thermal transport with cathodoluminescence thermometry

Kelly W. Mauser[1], Magdalena Sola-Garcia[1], Matthias Liebtrau[1], Benjamin Damilano[2], Pierre-Marie Coulon[3],

Stéphane Vézian[2], Philip Shields[3], Sophie Meuret[4], Albert Polman[1]

1. Center for Nanophotonics, NWO-Institute AMOLF, Amsterdam, Netherlands. 2. Université Côte d'Azur, CNRS, CRHEA, Nice, France. 3. University of Bath, Bath, United Kingdom. 4. CEMES/CNRS, Toulouse, France.

## ABSTRACT

Thermal properties have an outsized impact on efficiency and sensitivity of devices with nanoscale structures, such as in integrated electronic circuits. A number of thermal conductivity measurements for semiconductor nanostructures exist, but are hindered by the diffraction limit of light, the need for transducer layers, the slow-scan rate of probes, ultra-thin sample requirements, or extensive fabrication. Here, we overcome these limitations by extracting temperature from measurements of bandgap cathodoluminescence in GaN nanowires with spatial resolution limited by the electron cascade, and use this to determine thermal conductivities in the range of 19-68 W/m·K in three new ways. The electron beam acts simultaneously as a temperature probe and as a controlled delta-function-like heat source to measure thermal conductivities using steady-state methods, and we introduce a frequency-domain method using pulsed electron beam excitation. The different thermal conductivity measurements we explore agree within error where comparable. Our results provide novel methods of measuring thermal properties that allow for rapid, in-situ, high-resolution measurements of integrated circuits and semiconductor nanodevices, and open the door for electron-beam based nanoscale phonon transport studies.

**INTRODUCTION**

Increased attention must be paid to the effect of temperature on device performance in nanodevices due to highly concentrated energy densities and fewer heat conduction pathways through which waste heat energy can dissipate. Temperature control is crucial in many systems: nanowire single photon detectors must be cryogenically cooled to enter the superconducting regime and eliminate thermal noise[1]; nanowire lasers see a shift in lasing threshold and wavelength with temperature rises[2]; thermoelectric nanostructures rely on low thermal conductivity to generate large temperature gradients to increase efficiency of power generation or detection[3], and a microchip can have significantly varying gain and noise characteristics across its range of operating temperatures. Additionally, as integrated circuits shrink in size, the on-chip power density has increased by an order of magnitude over a decade, creating challenges in how to handle heat dissipation in nanoscale transistors[4]. This wide array of examples highlights the importance of careful thermal management designs in nanostructure devices to provide stable output and performance as these technologies increasingly move to smaller scales. Past research has studied engineering thermal conductivity by measuring or tailoring phonon mean free path spectra[5–13]. However, measuring both temperature and thermal conductivity of nanostructures is notoriously difficult.

A number of non-invasive methods have been devised to measure temperature and thermal conductivity on the nanoscale[14–17]. The highest spatial resolution thermometry methods include near-field scanning optical microscopy[18,19], scanning thermal microscopy[20,21], and transmission electron microscopy[22–24]. These methods can have a spatial resolution well below 100 nm but have generally slow data collection, cumbersome probes, or require very thin samples. Common thermal conductivity measurement methods include the 3ω method[25,26], the suspended microchip method[27–29], and time-domain thermoreflectance[8,9,13,30–33] and its variants[33]. These methods lack the spatial resolution of the high resolution thermometry methods listed above, require invasive or extensive fabrication, or a transducer layer, all of which hinders the ability to measure smaller domains or thermal boundary effects[31,33].

Here, we use a minimally explored nanoscale thermometry method, cathodoluminescence thermometry, which has spatial resolution limited only by the electron beam cascade size in the material. Cathodoluminescence (CL) is radiation emitted when a high-energy electron beam interacts with a material. In semiconductors, CL derives primarily from bandgap emission generated when high energy electrons lose energy to bulk plasmons through inelastic collisions, which then excite hot electron hole pairs which can thermalize and/or generate electron hole pairs that subsequently recombine by the emission of CL[34–36]. CL spectroscopy has been used in mineralogy[37], semiconductor characterization[38,39], and the study of plasmonic and photonic modes in metallic or dielectric nanostructures with nanometer resolution[40,41]. As we will show, at low beam currents, CL thermometry provides high-resolution non-invasive temperature measurements. At high beam currents, the beam acts like a nearly delta function heat source while simultaneously probing the temperature. While CL has been used previously as a thermometry technique[42,43], it has not, to our knowledge, been used for nanoscale thermal imaging or to study thermal conductivity.

We use the thermal bandgap shift in semiconductors to show first that we can map out the temperature profile from an electron-beam-induced heat source with the spatial resolution of a scanning electron microscope (SEM) electron cascade in a non-destructive manner, measuring temperature rises of over 500 K. Next, we use this thermometry technique to extract the thermal conductivity of GaN nanowires with three new, different methods; two methods using a DC electron beam current, and one method using a new technique involving an ultrafast electron beam blanker to provide an AC heating/thermometry source. The data obtained using the three methods compares favorably. With higher spatial resolution than state-of-the-art laser-based techniques[32] combined with fast scan speeds and no near-field probe or thin sample requirement, our work provides the means for a whole new realm of nanoscale phononic and thermal transport studies in semiconductors, including in-situ measurements of silicon integrated circuits.

**TEMPERATURE MEASUREMENTS**

The CL setup used for these measurements is shown in Fig. 1a and described further in previous work[44]. Briefly, a parabolic mirror inside of the SEM chamber with numerical aperture 1.46π sr is aligned over the sample so that the focal point of the mirror corresponds to the electron beam focus on the sample, and light is then collected and directed to a spectrometer. One of the GaN nanowires used is shown in Fig. 1b. The 200-300 nm diameter GaN nanowires were made by a top-down approach based on sublimation under vacuum (see Methods) and exhibit lasing properties[45] under optical pumping. We observe that the nanowires continue lasing while simultaneously undergoing high current electron beam irradiation indicating negligible degradation of the wires during CL measurements (see Supplementary Fig. 1 for more details). The nanowire was broken off the substrate on which it was grown and placed on the frame of a copper TEM grid covered with a 2 nm thickness of lacey carbon film, which thermally isolates the wire relatively well. Figure 1c shows a CL intensity map of the wire for emission between 360-400 nm, and the corresponding SEM image (collected during CL imaging) is shown in Fig. 1b taken at a beam current of 67 nA and electron energy of 5 keV. At each pixel in the CL map, the electron beam is focused at this point, and light is collected and the spectrum analyzed. Figures 1d-e illustrate a redshift in the peak bandgap emission of the CL spectrum as the electron beam becomes more centered on the wire, which we will explain shortly, and the colour of the spectra in Fig. 1d correspond to the location of the coloured dots on Fig. 1b-c. The small, non-red-shifted CL intensity observed for the beam placed next to the wire results from backscattered or secondary electrons from the substrate which excite CL from the wire and deposit little power inside. The most intense and most redshifted CL is observed for an electron beam centered on the wire. Using CASINO[46] Monte Carlo simulations, we determine that approximately 71% of the energy of the electrons is converted to heat in the wire. The remainder of the electron beam energy is mostly lost to backscattered and secondary electrons (Supplementary Fig. 2), with a negligible amount of energy lost to bandgap emission (Supplementary Fig. 3) and X-rays. Electron beam heating[47–50] calculations based on CASINO Monte Carlo simulations have previously been verified experimentally[47,48]. With a 67 nA electron beam, this corresponds to 238 µW of power being deposited in the wire in a nearly delta-function shaped power distribution (see Supplementary Fig. 2).

Our thermometry is carried out by tracking the shift in the peak bandgap emission energy as a function of temperature, due to thermal expansion of the lattice and changes in electron-phonon interactions with temperature[51,52]. As shown in Supplementary Fig. 4, we calibrate the wavelength shift with temperature in our GaN nanowires by measuring the bandgap shift as a function of temperature between 90-300 K using a 548 pA beam current in a liquid-nitrogen cooled cryogenic stage on our microscope. At a beam current this low the heating induced by the electron is negligible. Many of our measurements were carried out below room temperature within the range of our calibration curve to ensure our measurements were accurate; in some cases we extrapolate this curve to higher temperatures, following the Varshni phenomelogical formula[52],

$$E_g(T) = E_g(0) - \frac{\gamma T^2}{\beta + T} \qquad (1)$$

where $E_g$ is the bandgap as a function of temperature, $T$, $E_g(0)$ is the bandgap energy at 0 Kelvin (a fit parameter), and $\gamma$ and $\beta$ are constants[52]. The bandgap shift could alternatively be fit to an expression from O'Donnell and Chen[51]. While we focus on CL thermal measurements in GaN in this paper, a bandgap shift (red or blue) with increasing temperature can be seen in many other semiconductor materials[51,52]; CL spectra for intrinsic GaAs and p-doped Si wafers at different temperatures are shown in Supplementary Fig. 4 to demonstrate that the CL thermometry and the thermal conductivity measurement techniques presented here are not limited to use with GaN. Photoluminescence bandgap shifts in GaN nanowires have previously been used to measure temperature in a similar manner, but suffer from the poor resolution of the laser used as a heater/probe[53].

From our fit (Supplementary Fig. 4), we determined $E_g(0)$ = 3.471 eV (in GaN this corresponds to a donor-bound excitonic transition at low temperature, not the bandgap[54]), $\beta$ = 2609 K, and $\gamma$ = 2.25x10$^{-3}$ eV/K, which is similar to previous studies of GaN epilayers on sapphire substrates[54], with differences likely caused by different growth mechanism and the nanoscale geometries, and the fact that we fit a single Lorentzian to the entire near-band-edge PL spectrum to determine our effective bandgap instead of tracking shifts of individual exciton transitions the PL spectrum is comprised of. The root-mean-square error in temperature of our data around the line of best fit is 6.0 K. The thermal stage used had temperature accuracy of ± 1 K, and additional error likely comes from doping variations in the wires, which can be corrected for and will be discussed later. We fit the data in Fig. 1e with Equation 1 to create a temperature map (Fig. 1f) of the GaN nanowire resulting from electron beam heating at each pixel.

In order to reduce uncertainty in the thermal contact area between the wires and the substrate, nanowires were scattered over a copper TEM grid (Ted Pella G2000HA) with 6.5 μm diameter holes. Wires which straddled holes were heat sunk to the copper via electron beam assisted Pt deposition to fix the temperature at the ends of the wires during heating and reduce interfacial thermal resistance between the wires and the Cu TEM grid[28]. An SEM image of a wire in this configuration is shown in Fig. 2a. Following previous work[27–29,47–50,53,55,56], we treat the nanowires as 1D systems, and ignore thermal radiation and losses from CL (see Supplementary Fig. 3) in our analytical calculations. Finite element[57] simulations support this approximation. Figures 2b-e show temperature maps of the wire in Fig. 2a for different electron beam currents, in which the electron beam itself is used both as a heat source and as a thermometer. In these maps, the sample stage temperature was maintained at 161 K, and the peak CL wavelength was extracted for each pixel on the map and converted to temperature according to our calibration curve (Equation 1). Note that each pixel is measured when the electron beam is focused on that particular location. We use a 5 keV electron beam as Monte Carlo simulations indicate that at this energy most of the electron energy will be deposited within the wire (Supplementary Figure 2). A higher energy beam would give better spatial resolution but most of the electrons would pass through the wire without interacting, limiting heat deposition.

Several trends can be observed from the data in Fig. 2. First, in all images the largest temperature rise is observed toward the center of the wire, as expected theoretically for a 1D system with fixed temperature at both boundaries and an internal heat source. We can also see the high spatial resolution of the CL thermometry technique. Using higher beam currents we can generate temperature increases of over 200 K, showing the power of this technique to create temperature profiles from which the thermal conductivity can be derived, as we will show below.

### DC THERMAL CONDUCTIVITY MEASUREMENTS

We demonstrate three different methods to derive the thermal conductivity from the CL profiles, the first two being DC measurements, with analysis similar to Raman thermography or photoluminescence mapping found in other work[53,55,56]. In the DC measurement techniques, the wire is suspended over a hole in the TEM grid as shown in Fig. 2 and the inset of Fig. 4b and heated by a continuous electron beam and the steady-state temperature is extracted at every point along the wire as shown in Fig. 3a,d. We fit the temperature profile using theoretical models for 1D wires with temperature fixed by heat sinking with SEM-deposited Pt at both ends (the bridge method, Fig. 3c) or at one end (the slope method, Fig. 3f). In the bridge method, thermal contact resistance between the GaN and Cu TEM grid must be negligible[56], and in the slope method this thermal contact resistance is not important[55].

In the DC bridge method, both ends of the nanowire are heat sunk with SEM-deposited Pt and suspended over a bare copper TEM grid hole (see inset of Fig. 3a, Fig. 3c). We form an equivalent resistance model for the wire (described in more detail in Supplementary Note), shown in Fig. 3b, similar to previous work[56]. We assume there are two different thermal conductivities in the system: the thermal conductivity of GaN in the center of the wire, $\kappa_{GaN}$, and an effective thermal conductivity for a mixture of GaN and Pt closer to the Pt heat sinks, $\kappa_0$, attributed to the enlarged GaN nanowire radius due to excess Pt on the surface (see Supplementary Fig. 5, Fig. 3c). $L_1$ and $L_2$ demarcate the boundaries between the regions of different thermal conductivities and were treated as fit parameters. The system is represented in Fig. 3b by a thermal circuit model. Here, the thermal resistance is given by $R = lA/\kappa$, with $l$ being the relevant length of the particular segment and $\kappa$ the thermal conductivity of that segment. $l$ can change depending on the position of the heat source (see Fig. 3c, Supplementary Note), so the equations for the peak temperature rise, $\Delta T(x)$, as a function of $x$, the position of the electron beam heat source/thermometer, are

$$\Delta T(x) = \frac{\dot{Q}}{A}\left(\frac{\kappa_0}{x} + \frac{\kappa_0\kappa_{GaN}}{\kappa_{GaN}(L_1 - x + L - L_2) + \kappa_0(L_2 - L_1)}\right)^{-1}, \qquad 0 \leq x \leq L_1$$

$$\Delta T(x) = \frac{\dot{Q}}{A\kappa_0\kappa_{GaN}}\left(\frac{1}{\kappa_{GaN}L_1 + \kappa_0(x - L_1)} + \frac{1}{\kappa_0(L_2 - L_1) + \kappa_{GaN}(L - L_2)}\right)^{-1}, \qquad L_1 \leq x \leq L_2$$

$$\Delta T(x) = \frac{\dot{Q}}{A}\left(\frac{\kappa_0\kappa_{GaN}}{\kappa_{GaN}(L_1 + x - L_2) + \kappa_0(L_2 - L_1)} + \frac{\kappa_0}{L - x}\right)^{-1}, \qquad L_2 \leq x \leq L \quad (2)$$

where $\dot{Q}$ is the heat flux from the electron beam, $L$ is the total wire length, $A$ is the cross-sectional area of the wire as measured via SEM images, and $\Delta T(x) = T(x) - T_0$, where $T_0$ is the fixed temperature at $x = 0$ and $x = L$. Figure 3c shows this geometry in more detail. We fit temperature data obtained in Fig. 2e with Equation 2 in order to extract $\kappa_{GaN}$, $\kappa_0$, $L_1$, and $L_2$. This fit is shown with the data in Figure 3a. We find the thermal conductivity of the GaN region to be $\kappa_{GaN} = 22 \pm 4.7$ W/m·K, and the thermal conductivity of the edge region to be $\kappa_0 = 91 \pm 18.9$ W/m·K.

Several factors affect the accuracy of the determination of the parameters in the DC bridge method. First of all, our calibration curve only extends up to room temperature, while we extrapolate above room temperature in this analysis, creating some uncertainty. In future work this can be avoided by performing a more extended calibration. Second, a small variation in doping within each nanowire causes a ~1 nm variation in CL peak energy in different places along the wire (CL variation due to doping has also been observed previously in GaAs nanowires[39]), which also affects the temperature calibration. This could be corrected for by using, as a reference, low-current CL measurements that probe the bandgap at each position, as we do later. Here we use a relatively high beam current to create a fairly high temperature rise to more effectively smooth out the 1 nm variations in CL peak shift along the wire (since spectral peak shifts in this measurement are much larger than 1 nm). Because we heat sink both ends of the wire, a relatively high current is needed to achieve a large redshift. We note that in most of our nanowires a 1-2 μm region at one end shows both less CL intensity (see Fig. 1c) and a slightly blue-shifted CL peak relative to the rest of the nanowire (measured at low electron beam currents), while towards the other end of the wire an abrupt increase in CL counts with a slight red-shift is observed. This is due to the doping profile introduced during nanowire growth (see Methods). We note that in the DC bridge model we neglect the interfacial thermal resistance[56] between the GaN and the Pt/Cu at either end as it is small compared to the thermal resistance of GaN, which we verify with our measurements by ensuring temperature rises are very small near the Cu heat sinks, as seen in Fig. 3a. The interfacial thermal resistance is not always negligible between the end of the wire and the Cu substrate, which we observe in our measurements as a discontinuity between the temperature of the nanowire near the Cu and the Cu temperature, known from a thermometer in the sample stage (within 1 K accuracy) to which the Cu is thermally connected with silver paint. To overcome this, a different method can be used to measure thermal conductivity, the DC slope method[55].

In the DC slope method, only one end of the wire is heat sunk (see inset of Fig. 4b) and the other end extends into the center of the hole. In this method[55], the temperature rise when the electron beam is at position $x$ away from the edge of the hole (Fig. 3f) is given by

$$\Delta T(x) = \left(R_c + \frac{x}{A\kappa_{GaN}}\right)\dot{Q}, \qquad (3)$$

where $R_c$ is thermal contact resistance between the wire and the Cu frame. If we find the slope, $s$, of this line, $d\Delta T/dx$, and solve for $\kappa_{GaN}$, we get the expression $\kappa_{GaN} = \dot{Q}/(sA)$. We determine $A$ (wire cross-sectional area) from SEM images. $\dot{Q}$ (heat flux) we determine from the measured electron beam current correcting for energy lost to backscattered or secondary electrons (determined from CASINO Monte Carlo simulations). In the case of the "o" data points in Fig. 3d, we also correct for larger electron beam sizes which resulted as a consequence of using large currents. The slope is found by fitting a line to the temperature profile of the wire sufficiently far from the Pt contacts to avoid the effect of the Pt/GaN thermal conductivity seen in Fig. 3a. We additionally subtract the doping profile (resulting from variations in intentional Si-doping during growth) of the wires found under low electron beam current[39] to correct for the 1 nm doping variations along the wire as discussed above (see Supplementary Figure 6, Methods). The profiles of two such wires are shown in Fig. 3d with thermal conductivities specified in the figure caption, ranging from $30 - 66$ W/m·K with errors ranging from 10-21% which derive primarily from uncertainty in $A$ due to small variations in diameter along the length of the wire, and in the case of the "o" data points, from a 10% error in $\dot{Q}$, as discussed below.

The benefit of the DC slope method over the DC bridge method lies in the ability to neglect thermal contact resistances. Additionally, larger temperatures can generally be reached in wires only thermally connected on one end. Overall, DC methods suffer from strong dependence on localized doping variations. This can be overcome by extracting bandgap variations due to doping profiles with low electron beam currents as was done in Fig. 3d. There is additional uncertainty that comes from the heat flux in the wire, in the case of Fig. 3a and "o" data points in Fig. 3d. Because large currents are needed to raise temperatures for good signal-to-noise ratio, larger apertures must be used in the electron column which leads to larger spot sizes[58]. This is generally negligible in comparison to the size of the electron cascade, unless the aperture is removed entirely and less of the incident electron beam impinges upon the wire, adding some uncertainty to the measurements of the heat flux $\dot{Q}$. In the measurements of Fig. 3a and Fig. 3d ("o" data points only), by examining the loss of resolution in secondary electron images as a result of increased electron beam size, we calculate that only approximately 20-50% of the electron beam is reaching the nanowire without an aperture. Thus, the current actually reaching the nanowires was 64.0 nA for Fig. 3a, and for Fig. 3d Wires A and B "o" data points, 15.7 nA and 11.0 nA, respectively. To double-check the veracity of our thermal conductivity measurements using the DC slope method, an aperture was used when collecting the "x" data points in Fig. 3d, leading to less current (5.6 nA and 3.2 nA for Wires A and B, respectively), a smaller temperature rise in the wire, but all of the measured electron beam current striking the wire in a several nm-sized spot. The dependence on knowing $\dot{Q}$ to a high degree of accuracy can be overcome by using AC methods to extract thermal conductivity, as discussed in the next session.

**AC THERMAL CONDUCTIVITY MEASUREMENTS**

In the AC thermal conductivity measurement technique, the column of the SEM was equipped[59] with a high-frequency electrostatic beam blanker to modulate the electron current in a square wave on/off pattern. The sample configuration is the same as that used in the DC slope method described above, in which one end of the nanowire is heat sunk with SEM-deposited Pt, and the other end is free (Fig. 4b, inset, Fig. 3d).  In this method, we focus the electron beam on the free end of the wire for the duration of the experiment and vary the electron beam current frequency with a waveform generator between 100 Hz and 5 MHz (Supplementary Fig. 7).  Data collection for the studied frequency range took several minutes total.   Solving the 1D time-dependent heat equation (using one Dirichlet and one time-dependent-periodic Neumann boundary condition) for the quasi-steady state temperature (after all transients have subsided) at the free end of the nanowire, temperature varies according to the expression (see Supplementary Note for more details)

$$T(\omega, t) = T_0 + \frac{4\dot{Q}L}{A\kappa\pi} + Re\left\{ \sum_{m=1,3,5,\ldots}^{\infty} \frac{4\dot{Q}x}{A\kappa\pi m i} \frac{\tanh\left(\sqrt{\frac{\omega L^2 C_p \rho m}{2\kappa}}(1+i)\right)}{\sqrt{\frac{\omega L^2 C_p \rho m}{2\kappa}}(1+i)} e^{im\omega t} \right\}, \quad (4)$$

where $T_0$ is the temperature of the fixed end/Cu frame, $A$ is wire cross-sectional area, $\kappa$ is thermal conductivity (we assume uniform thermal conductivity in the wire), $\rho$ is density of GaN[60] (6150 kg/m$^3$), $C_p$ is heat capacity[60] (490 J/kg·K), and $L$ is wire length starting from the edge of the Pt deposition. Because we use a spectrometer with a long exposure time (40 ms or longer) compared to the modulation frequency of the beam, we measure the average temperature over the half period when the electron beam is on (Fig. 4a inset),

$$\bar{T}_{meas}(\omega) = T_0 + \frac{4\dot{Q}L}{A\kappa\pi} + \frac{8\dot{Q}L}{A\kappa\pi^2} Re\left\{ \sum_{m=1,3,5,\ldots}^{\infty} \frac{1}{m^2} \frac{\tanh\left(\sqrt{\frac{\omega L^2 C_p \rho m}{2\kappa}}(1+i)\right)}{\sqrt{\frac{\omega L^2 C_p \rho m}{2\kappa}}(1+i)} \right\}. \quad (5)$$

The sum arises from the Fourier decomposition of a square wave.  We find that at low modulation frequencies (e.g. 100 Hz) the time-averaged temperature of the GaN wire is higher, and therefore it has a more red-shifted CL spectrum than at higher frequencies (e.g. 5 MHz) as shown in Fig. 4a.

Wavelength spectra for each modulation frequency were fit to extract the temperature data shown for two different wires in Fig. 4b, the same wires of which were analyzed in Fig. 3d using the DC slope method. Several different frequency sweeps were performed for each wire at slightly different locations at the end of the wires, which correspond to the different fit curves in Fig. 4b. The solid lines are fits corresponding to Equation 5. The error in thermal conductivity due to the variation in each curve in Fig. 4b is approximately 4.1% and 6.6% for Wires A and B, respectively. The mean value of $L$ was taken from data from the DC slope method (i.e. where a kink in slope of temperature versus x appears indicating Pt deposition, not shown in Fig. 3d but visible in Fig. 3a), and uncertainty was measured as the range of the visible thickening of the nanowire radius due to leakage Pt deposition, as seen in SEM images. Uncertainty in $L$ ranges from 5-10% for the wires. This leads to an uncertainty in thermal conductivity due only to contributions from $L$ uncertainty of 20% and 7.5% for Wires A and B, respectively. Thus, uncertainty in the AC method thermal conductivity measurements has the largest contribution from $L$ uncertainty. The leakage of Pt deposition onto the wires was the primary culprit for this uncertainty, as it is unclear where the exact location of the "fixed" temperature end of the wires is. The uncertainties can be strongly reduced by further control over the sample geometry. The DC slope method does not rely on knowledge of $L$ to extract thermal conductivity, and the DC bridge method treats the equivalent of $L$ as a fit parameter, incorporating error into the fit model.

The benefit of the AC method is that the value of the electron beam heat flux, $\dot{Q}$, does not need to be known in order to extract thermal conductivity, unlike in the DC methods, where these parameters are 100% correlated. As $\dot{Q}$ is determined by using a combination of measurement of the electron beam current and CASINO and Monte Carlo simulations, and can be heavily influenced by electron beam shape (in the case of unapertured electron beams, as discussed above), uncertainties in $\dot{Q}$ are a significant source of error in the analysis. By using a spectrometer to average the frequency-dependent optical response in time, we are summing non-negligible shot noise (from electrons striking the nanowires and from generation/recombination of carriers in the GaN nanowires) over a wide electrical bandwidth. We could further shrink the uncertainty in thermal conductivity in these AC measurements by using a bandpass filter to isolate a small wavelength range near the bandgap CL emission and monitor amplitude modulation in this band via lock-in detection during pulsed electron beam excitation, thereby drastically improving signal to noise ratio as the bandwidth over which noise is summed with lock-in methods is small.

**COMPARISON OF THERMAL CONDUCTIVITY METHODS**

The thermal conductivity of bulk GaN at room temperature reported in the literature is fairly high at 130-220 W/m·K[61,62]. GaN nanowires previously studied with the suspended microchip method or with photoluminescence have reported thermal conductivity values of 13-19 W/m·K[29] for smaller diameter nanowires and <80 W/m·K[53] for wires of similar diameter to those we studied. Therefore, the thermal conductivities of 22-68 W/m·K found in this study are reasonable. Several studies attributed the deviation from the bulk values to decreased phonon mean free path due to large mass-difference scattering from Si-impurities[29,63,64]. One study found additionally that boundary scattering, phonon confinement, and the change in nonequilibrium photon distribution significantly contributed to the decrease in thermal conductivity in nanowires when compared to bulk[64]. Si-impurities are present in our nanowires, as the nanowires were intentional doped with Si during growth (see Methods). The nanoscale thermometry presented here provides avenues for detailed studies of heat flow in confined geometries.

The data shown in Fig. 3d are from the same wires as the data shown in Fig. 4b, allowing for direct comparison between the DC slope and AC methods. The extracted thermal conductivities are within error of each other for Wires A and B. Both the DC and AC methods have their advantages and disadvantages. The DC methods allow for easier examination of the heat-sinking quality at the boundaries of the materials during data collection, and have higher spatial resolution than the AC method (in our microscope).  Additionally, drift during measurements is easier to spot and correct for with the DC method.  The DC method is also more sensitive to variations in doping in the wire.  With extra large temperature rises, as we saw in the wire used in the DC bridge method, the doping variations played less of a role in determining thermal conductivity as temperature rises caused much larger bandgap shifts than doping variations did.  In the DC slope method with smaller temperature rises, we had to correct for variations in the bandgap emission due to doping (on the order of 1 nm) in the nanowire by extracting bandgap variations along the wire found with low electron beam current.  On the other hand, knowledge of the thermal contact resistance was not necessary in the DC slope method, whereas it was critical in the DC bridge method.  In fitting the DC data, thermal conductivity is 100% correlated with both electron beam heat flux and wire cross-sectional area, both of which had the largest sources of uncertainties.

The AC method, as it relies on the electron beam being focused on a single point on the wire, can be collected much faster.  In our particular SEM electron column, spatial resolution is worse in the AC method due to changes in the electron beam optics necessary to place the beam into conjugate mode with the focus of the beam between the two blanking plates[59].  Alternative microscope column designs with optimized beam cross-over, which are available commercially as well, will resolve this issue. In all cases, we assume that thermal conductivity is constant along the wire, and not affected by doping, however, it has been shown thermal conductivity of GaN can decrease with increased doping concentration[63] (see Supplementary Fig. 6 for discussion).  Accuracy could suffer if thermal conductivity is not uniform throughout the wire as the mathematical model does not account for this (see Supplementary Figs. 6, 8).  In fitting the data, thermal conductivity can be extracted accurately without knowledge of both electron beam heat flux and wire cross-sectional area in the AC method, in contrast to DC methods. The AC method may also suffer less from carrier accumulation in the bandgap, such as blue-shifting caused by the Burnstein-Moss effect[65,66].  We did not see such an effect in our wires, but it could play more of a role in other materials. In all cases, we assumed heat capacity and density were constant with temperature.

**DISCUSSION**

We have presented a method of cathodoluminescence nanothermometry for semiconductors along with three different methods for using this thermometry method to measure thermal conductivity of GaN (or other semiconductor) nanowires. CL thermometry can be used with very low currents in order to measure temperatures *in-situ* without heating the sample, or can be used with high currents to act additionally as a delta-function-like heat source to study thermal transport. We additionally showed that alongside GaN, both Si and GaAs exhibit shifts in CL bandgap emission with temperature, indicating that the temperature mapping and thermal transport measurements examined here are broadly applicable to other semiconductors and could find uses in examining integrated circuits in-situ to find defects, for example. The thermal conductivity measurement methods explored here are fairly rapid and have low fabrication requirements. The existing framework for laser-based pump-probe measurements of thermal conductivity, like time-domain thermo-reflectance, could easily be translated into the SEM using CL nanothermometry which could result in 100x better spatial resolution than these state-of-the-art methods[32]. Because of the high resolution, high scan speeds, and high level of control an SEM offers, CL nanothermometry-based methods offer an enticing framework in which to study phonon dynamics, ballistic transport, and near-field heat transport phenomena hitherto unmeasurable.

**Methods**

*Nanowire fabrication*

The nanowire fabrication process using a top-down approach combining displacement Talbot lithography and selective area sublimation has been detailed previously[45]. A GaN layer was grown by metalorganic-vapour phase epitaxy on c-plane (0001) sapphire substrates starting with a 2 µm undoped GaN layer followed by 5 µm of Si-doped ($5 \times 10^{18}$ cm$^{-3}$) GaN.   A 60 nm-thick Si$_x$N$_y$ deposited by plasma enhanced chemical vapor deposition on top of the GaN layer was patterned with displacement Talbot lithography to get a hexagonal array of nanodisks (diameter of 515 nm) with a pitch of 1.5 µm[67].  The sample then underwent selective area sublimation between 900 and 920°C in a molecular beam epitaxy chamber for 8 hours in order to define the 7 µm nanowires.  The Si$_x$N$_y$ mask was etched using a HF-based solution. Some wires shown in this manuscript are slightly shorter than 7 µm due to breaking when separated from the growth substrate or variations in the sublimation rate due to temperature gradient between the center and the edge of the wafer.

## Author Contributions

K.M., A.P., and S.M. conceived the experiment, B.D., P.C., S.V., and P.S. fabricated the nanowires, K.M. measured the data, K.M. analyzed the data, S.M., M.S.G., M.L., and K.M. built the measurement systems, all authors contributed to the manuscript.

## Acknowledgements

The authors would like to thank V. Neder and I. Hoogsteder for assistance with electron-assisted Pt deposition. This work is part of the research program of AMOLF which is partly financed by the Dutch Research Council (NWO). This project has received funding from the European Research Council (ERC) under the European Union's Horizon 2020 research and innovation programme (grant agreement No. 695343).  This work has been supported by the French National Research Agency (ANR) through the project NAPOLI (ANR-18-CE24-0022).

Competing financial interest: A.P. is co-founder and co-owner of Delmic BV, a company that produces the cathodoluminescence system that was used in this work.

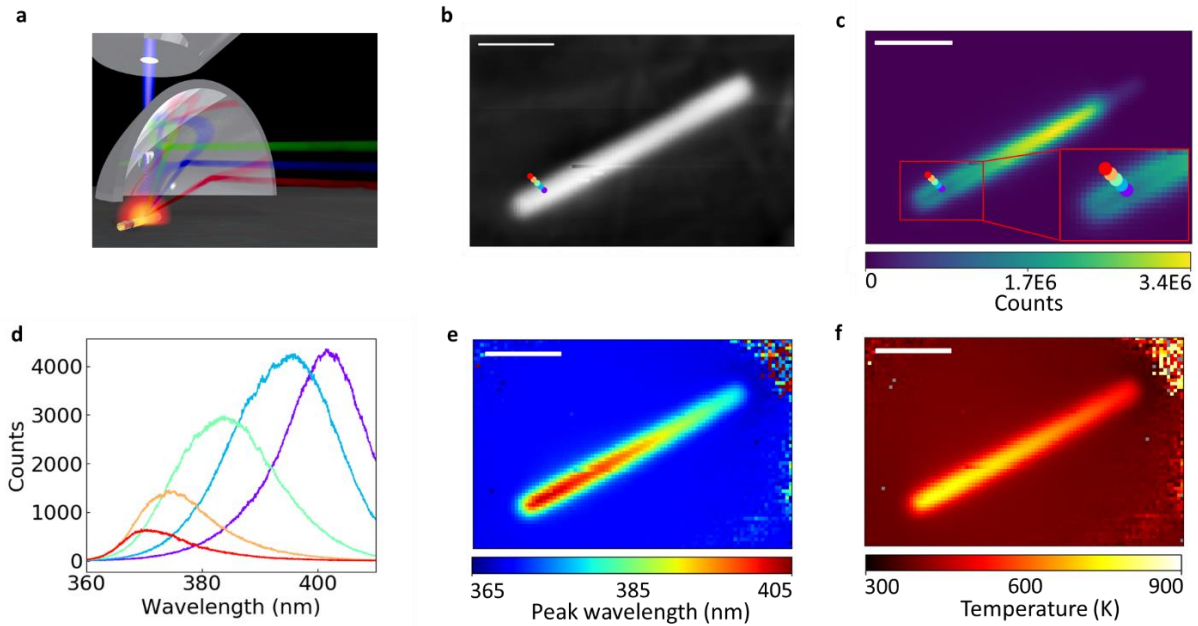

**Figure 1 | Nanoscale thermometry CL measurement apparatus and monitored signals. a**, Schematic of cathodoluminescence (CL) measurements on a semiconductor nanowire. An electron beam heats/excites the semiconductor nanowire and incoherent CL is collected by the high-numerical-aperture parabolic mirror and directed into a spectrometer. **b**, SEM image of a GaN nanowire. **c**, CL counts integrated between the wavelengths of 360-400 nm. (Inset) Zoomed-in region of GaN wire. **d**, CL spectra. Each spectrum is obtained at the position of the corresponding colour dot in both **c** and **d**. The amount of spatial overlap of the electron beam and the nanowire dictates the energy absorbed in the nanowire from the electron beam, resulting, when the beam is centered on the nanowire, in a maximum temperature rise and corresponding redshift of the CL emission according to Equation 1. **e**, Peak CL wavelength extracted by fitting the spectra corresponding to each pixel with a Lorentzian. **f**, Temperature map measured when the electron beam is focused at each pixel, obtained by fitting the data in **e** to Equation 1. Scale bars are 1 µm.

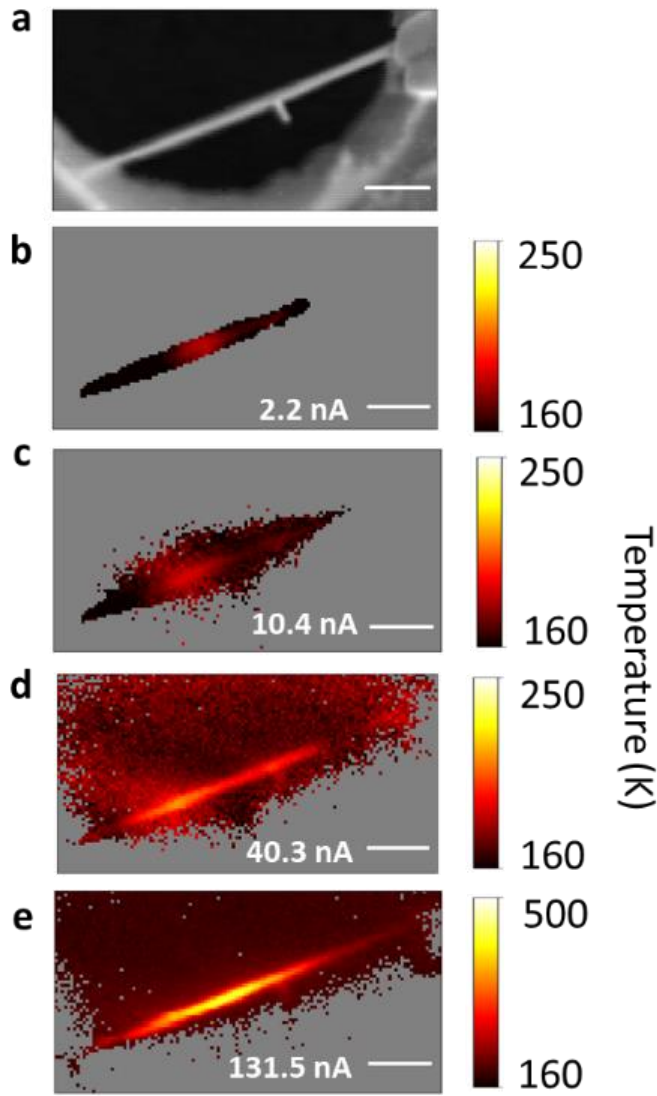

**Figure 2 | Nanoscale temperature measurements at variable electron beam currents. a**, SEM of suspended GaN nanowire with Pt heat sinks on either end. **b-e**, Temperature measurements of GaN nanowire in **a** at the specified currents. Gray regions indicate pixels which did not exhibit a peak in the CL spectrum above 100 counts and 1 nm in width or which could not be fit. Scalebars are 1 μm. The base temperature in all measurements is 161 K.

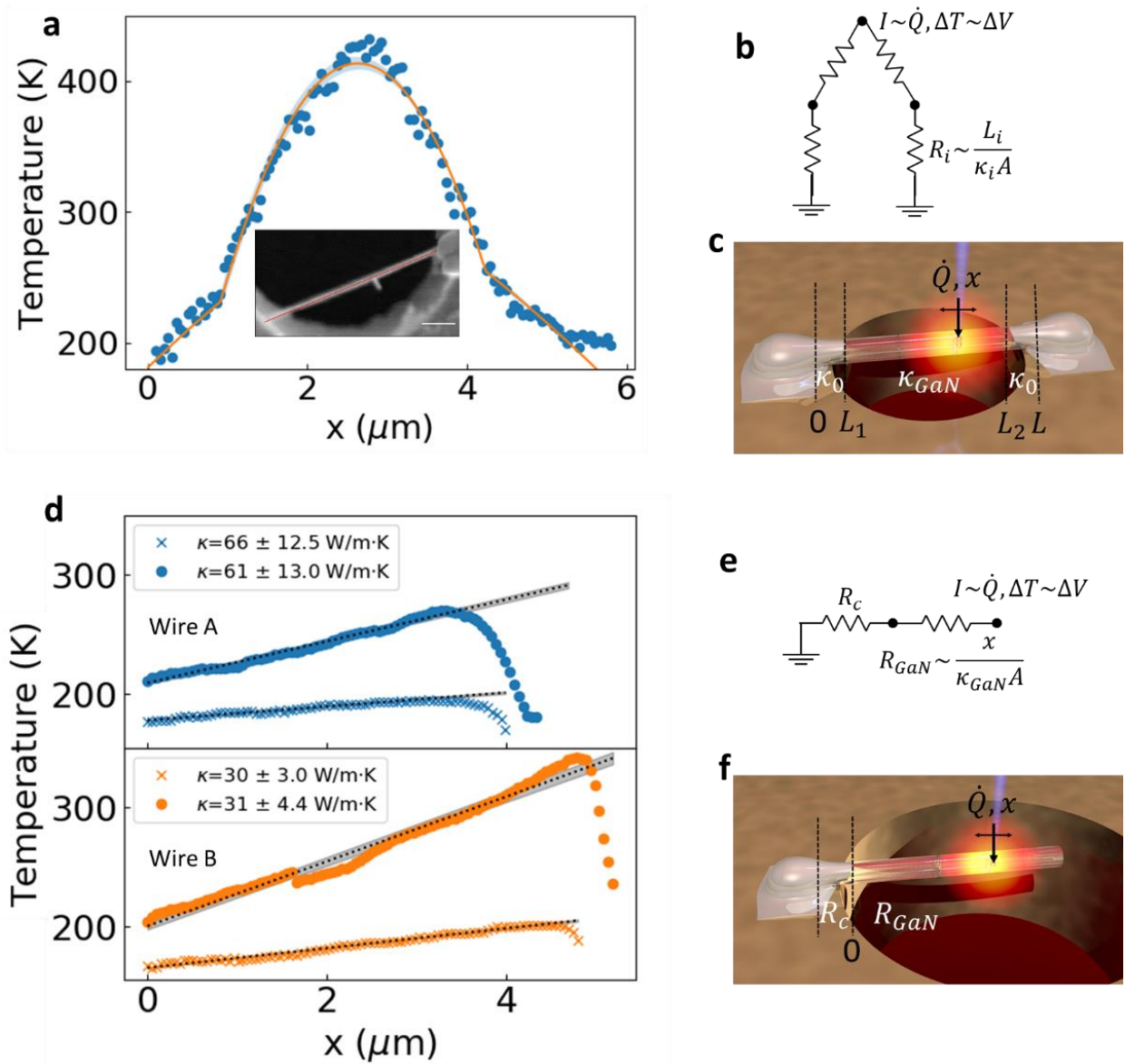

**Figure 3 | Probing nanowire thermal conductivity with DC electron beam. a**, Measured temperature as a function of position along the cut through the GaN wire in Fig. 2a shown in the inset. Orange line is best fit to data using Equation 2 (DC bridge method), and blue shading is ± one standard deviation of the fit error. We find a thermal conductivity of the GaN nanowire of 22 ± 4.7 W/m·K, and of the Pt/GaN portion 91 ± 18.9 W/m·K. Base temperature for these measurements is 161 K. Wire radius is 118 nm. **b**, Thermal circuit model for the DC bridge method, shown here for the case of $L_1 \leq x \leq L_2$, where $x$ is the location of the electron beam (see Supplementary

Note for more details). **c**, Schematic of temperature profile in the wire corresponding to the DC bridge method and values in Equation 2. **d**, Demonstration of DC slope method for determining thermal conductivity of two different nanowires with fixed temperature at one end (see example wire in inset of Fig. 4b). "x" data points are from apertured electron beams with nm spot sizes. The "o" data points are from data collected with unapertured electron beams, which result in a less well-defined spot size. The corresponding thermal conductivities are shown in the legends. Radius of the nanowires is 130 nm and 123 nm for Wires A and B, respectively. **e**, Thermal circuit model for the DC slope method. **c**, Schematic of temperature profile in the wire corresponding to the DC slope method and values in Equation 3.

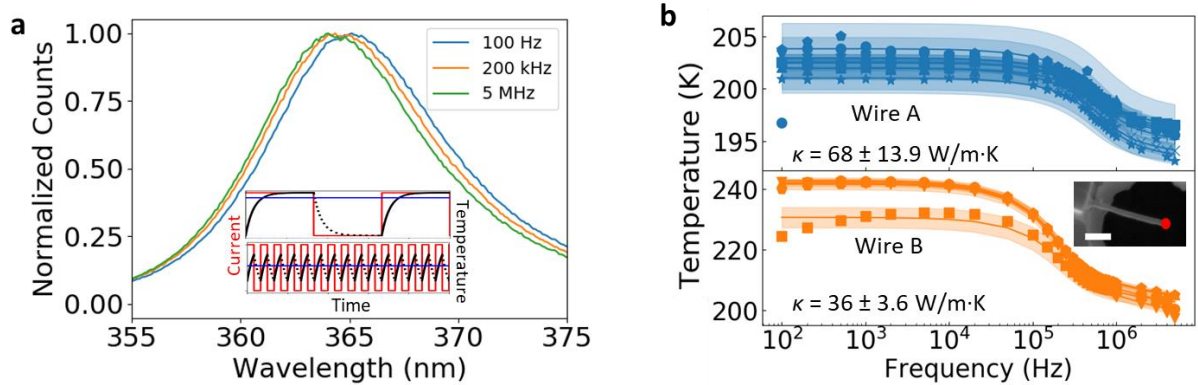

**Figure 4 | Cathodoluminescence thermal conductivity measurements in the frequency domain. a,** Cathodoluminescence (CL) spectra as a function of wavelength for a 100 Hz (blue), 200 kHz (orange), and 5 MHz (green) square wave electron beam excitation current. (inset) CL is only emitted when electron current (red) is flowing, so the temperature read by CL (black, solid line) will be on average (blue) higher for lower frequencies. The electron beam current used in this measurement to heat/probe the nanowire was 42 nA DC (the current measurement was taken without modulation, with modulation the DC measured current is half that value). **b,** Temperature as a function of electron beam square wave frequency for the same wires from Figure 3d with one end held at a fixed temperature. Each plot shows several different data collection runs (represented by different marker types) for the same wire at slightly different locations on the end of the wire. The solid line is the best fit line to the data, and the shaded regions are +/- one standard deviation of error in the fitting function. The extracted thermal conductivities are shown in each plot. Error is a combination of standard deviation of thermal conductivity extracted from plot data and percent error in measurement due to uncertainty in length measurements. The electron beam current used in this measurement to heat/probe the nanowire was 10 nA and 14 nA DC for Wires A and B, respectively (the current measurement was taken without modulation, with modulation the DC measured current is half the given value). (inset) SEM image of wire similar to the ones CL data were collected from. Scale bar is 2 μm.

# Probing nanoscale thermal transport with cathodoluminescence thermometry

Kelly W. Mauser[1], Magdalena Sola-Garcia[1], Matthias Liebtrau[1], Benjamin Damilano[2], Pierre-Marie Coulon[3], Stéphane Vézian[2], Philip Shields[3], Sophie Meuret[4], Albert Polman[1]

1. Center for Nanophotonics, NWO-Institute AMOLF, Amsterdam, Netherlands. 2. Université Côte d'Azur, CNRS, CRHEA, Nice, France. 3. University of Bath, Bath, United Kingdom. 4. CEMES/CNRS, Toulouse, France.

**Supplementary Information**

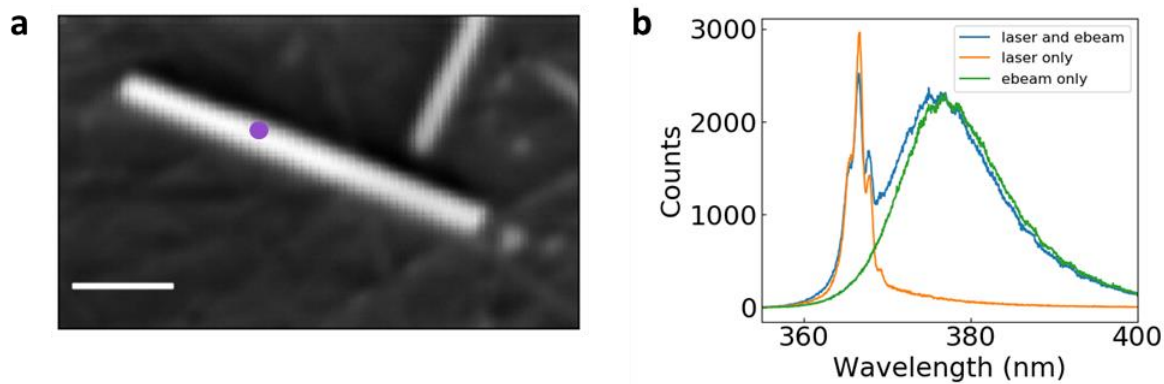

**Figure S1 | Optically-induced lasing during high current electron beam irradiation. a**. SEM image of GaN nanowire. The shown wire exhibits optical-induced lasing when pumped. The purple dot corresponds to one pixel, with the corresponding CL spectrum shown in **b**. Scale bar, 1 μm. **b**. Optically induced lasing spectrum (orange, 250 fs, 258 nm laser pulsed at 202 kHz with 1.8 mW average power) from the nanowire in **a**, CL spectrum (green, 5 kV, 79 nA) at the location indicated in **a**, and spectrum from simultaneous electron beam and laser illumination yielding both CL and optically-induced lasing (blue). The wire showed no noticeable degradation in lasing for over an hour while the electron beam was repeatedly scanned over the wire, indicating that the electrons are not destroying the lasing properties of the wire in a noticeable way. We do observe that the CL yield of the GaN wires does decrease slightly with time after scanning a wire ~30 times or more. We attribute this to electrons filling trap states in GaN in addition to carbon deposition on the surface of the nanowire which increases the surface recombination velocity[1,2]. Whether or not an electron beam damages GaN has been studied previously[1,2]. It has been determined that at the low energies found in an SEM (<30 keV), defects in the atomic lattice are not generated[3]. However, the electron beam can activate existing defects such as Ga vacancies. This, along with charged surface states, can increase the non-radiative recombination rate, yielding a CL intensity that decreases with time until it reaches a steady state when all defects/states have been activated. Carbon deposition also occurs in an SEM, and while not thick enough to optically

block CL, its presence has been found to significantly enhance the surface recombination velocity of GaN.

These effects occur at all electron beam currents and are semi-reversible.

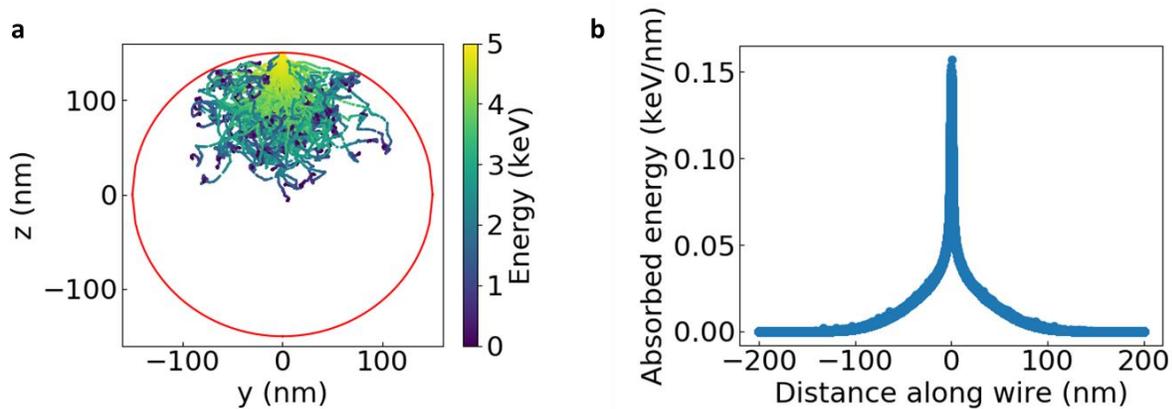

**Figure S2 | CASINO Monte Carlo simulations of GaN nanowire. a,** Trajectory and energy of 5 keV

primary electrons in SEM as calculated by CASINO Monte Carlo simulations in a 150 nm radius GaN wire,

outlined in red. **b,** Absorbed energy along the long axis of the nanowires as a function of wire length,

summed over the cross-sectional area of the wire at point and normalized by number of electrons. The

GaN density is 6.15 g/cm$^3$ and 100,000 5 keV electrons were simulated with a 5 nm incident electron beam diameter. CASINO simulations compute energy lost by an electron while undergoing collisions.

We assume energy lost by electrons in the simulation is converted into heat in order to calculate absorbed energy[4].

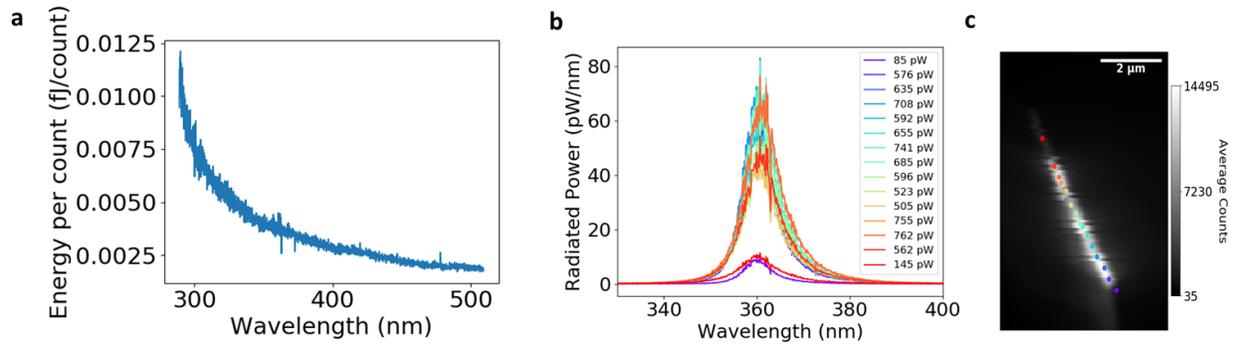

**Figure S3 | Radiated power from cathodoluminescence as a function of wavelength. a**, Calibration of detector for single crystal aluminum crystal based on a previous method[5]. A theoretical calculation[6] of the energy per unit bandwidth produced by transition radiation from Al per electron was divided by experimental counts from our detector to extract the curve shown. **b**, Radiated power spectrum for different pixels with locations corresponding to the dots in **c**. Legend gives the total integrated power for each pixel. Power was calibrated using the curve in **a**. Slight variations in the system alignment can lead to approximately a 30% change in radiated power in our system[5], therefore the radiated power shown here will have an error of about 30%. As the power deposited in the wire by the electron beam is on the order of hundreds of μWs, radiated power loss from cathodoluminescence is negligible. **c**, CL intensity map with colored dots corresponding to the location of the pixel where the spectra in **b** were taken.

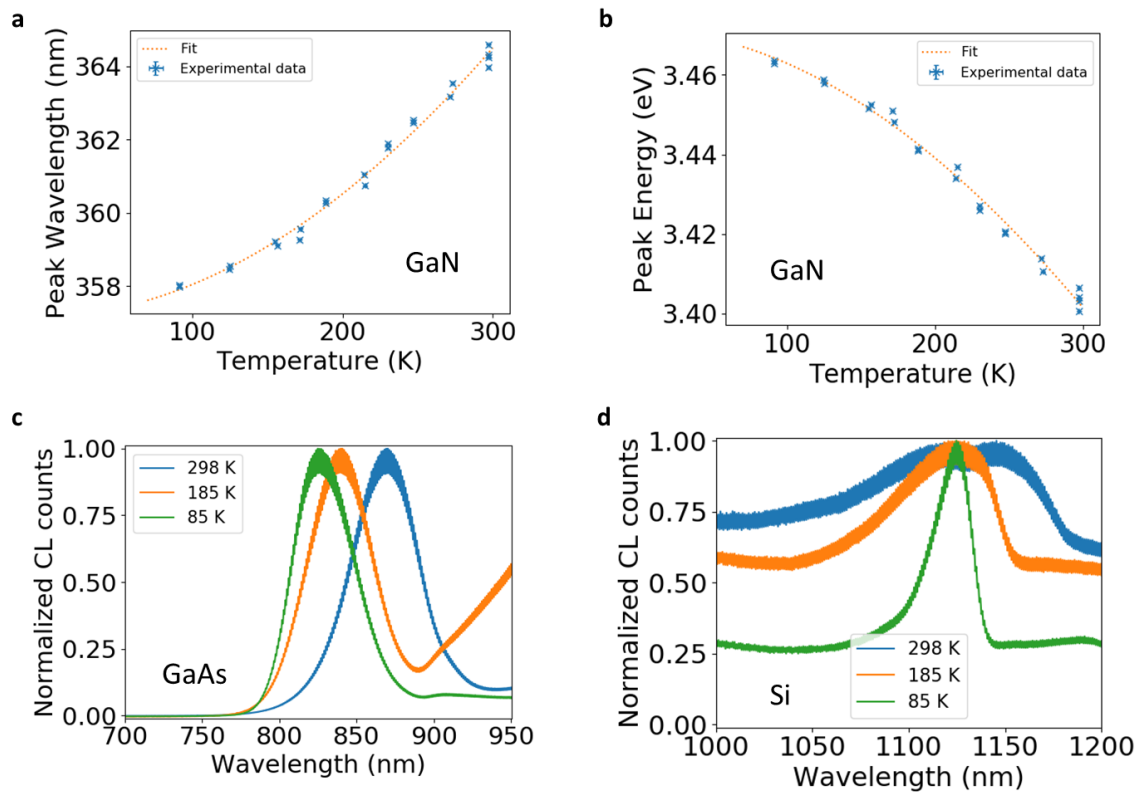

**Figure S4 | Cathodoluminescence spectral shifts with temperature. a,b,** Calibrated CL peak wavelength and energy versus temperature in GaN nanowires scattered on a Si substrate. The electron energy is 5 keV and electron beam current is 548 pA to minimize heating of the nanowires. The CL spectra were fit with a Lorentzian (for simplicity) to determine the peak, and the fit curve is from Equation 1, with constants $E_g(0)$ = 3.471 eV, $\gamma$ = 2.25×10$^3$ eV/K, and $\beta$ = 2609 K. The root mean square error of the data around the line of best fit is 6.0 K. The thermal stage used had temperature accuracy of ± 1 K, and additional error likely comes from doping variations in the wires. **c,** Normalized CL spectra at different temperatures for an intrinsic GaAs wafer from 5 keV electrons. **d,** Normalized CL spectra at different temperatures for a p-doped Si (5-10 Ω·cm) wafer from 30 keV electrons. GaAs spectra can likely be fit with a Lorentzian in the same manner as GaN to estimate the bandgap shifts, but due to the extensive broadening of the Si peak in addition to red-shifting of the bandgap, a Voigt or other asymmetric function may be needed to fit the spectra to identify the bandgap shifts. In all plots, the sample is

physically and thermally adhered to the SEM thermal stage with silver paste, and the sample temperature is controlled via the thermal stage temperature.

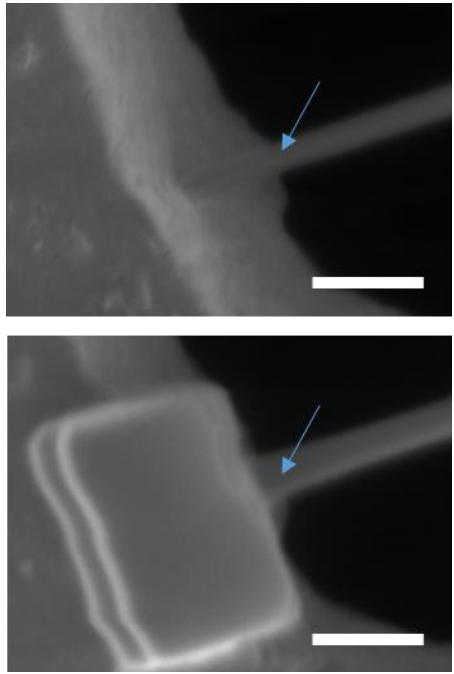

**Figure S5 | Pt deposition on edges of wires.** SEM of nanowire before (top) and after (bottom) Pt was deposited using focused electron-beam-induced deposition. Thickening of the wire near the Pt deposition is apparent (see arrows). Scalebars are 1 μm.

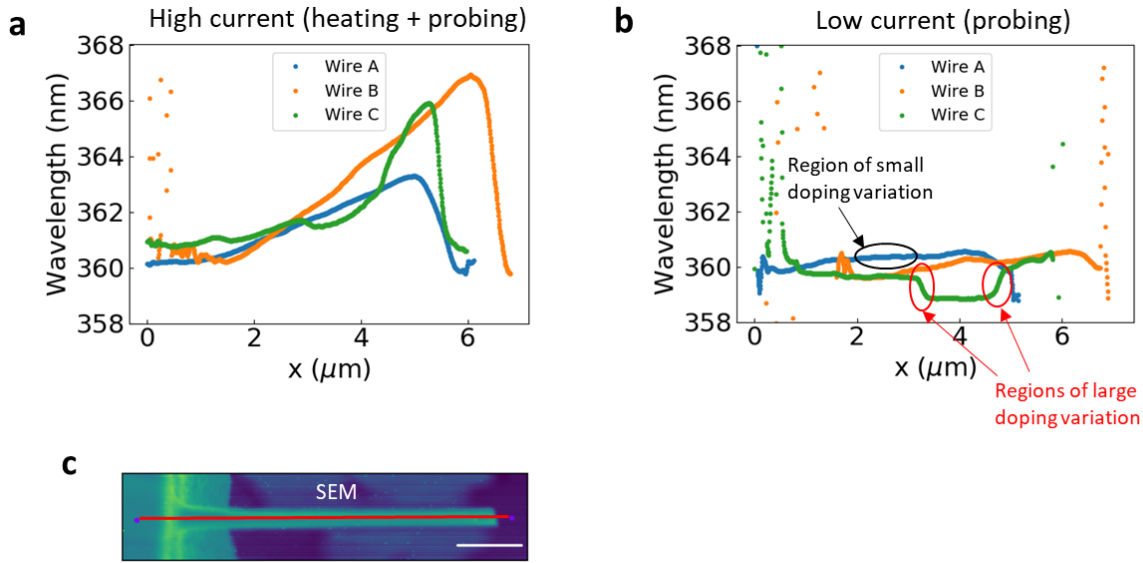

**Figure S6 | Influence of doping on peak CL wavelength. a**, Uncorrected peak wavelength data corresponding to Wires A (blue), B (orange), and C (green) with temperature data in Fig. 3d and Supplementary Figure 8, along the line shown in the SEM in **c** for a representative wire. The wires were irradiated with 5 keV electrons with beam currents of 15.6, 11.0 and 9.3 nA for Wires A, B, and C, respectively, leading to heating in the wire which causes the redshift in CL emission peak. **b**, The same wires measured in **a** are shown measured with electron beam currents of around 1 nA which negligibly heat the wires. Variations in peak wavelength here are presumably caused by variations in doping concentration in the nanowires, as was observed in other nanowires[7]. Red circled regions show locations of large doping variations in Wire C (green), while the black circled region shows a region of relatively little doping variation in Wire A (blue). The large doping variations in Wire C could be responsible for differences in thermal conductivities measured using the DC slope method and AC method and shown in Supplementary Fig. 8. It has been shown that the thermal conductivity of GaN can decrease with increased doping concentration[8]. That study[8] showed a factor of ~2 or more decrease in thermal conductivity between undoped GaN carrier concentrations (~$10^{17}$ cm$^{-1}$) and the carrier

concentrations found in our nanowires (~$10^{18}$ cm$^{-1}$).  We know that the nanowires used in our study have 2 um of undoped GaN at one end, which we can observe in low-current (non-heating) CL measurements as an approximately 1 nm blue shift in bandgap emission peak and a decrease in CL intensity.  For nanowires A, B, and the nanowire used in Fig. 3a, this undoped end of the GaN nanowires was the end covered in Pt and heat sunk to the edge of the Cu membrane, leaving the doped portion (with relatively minor doping variations as shown in **b**) exposed and probed with the electron beam.  Wire C, as seen in **b**, has larger doping variations and so could have variations in thermal conductivity along its length, invalidating our models which assume uniform thermal conductivity.    In both **a** and **b**, peaks and dips of data points at either end of the wire are due to low CL counts and inability of the fitting function to find the peak wavelength.  To correct for doping for the DC slope analysis in Fig. 3d and Supplementary Fig. 8, an average peak wavelength emission was chosen for a low current cross-cut and deviations from this average were subtracted from the high current cross-cut.  Scale bars, 1 μm.

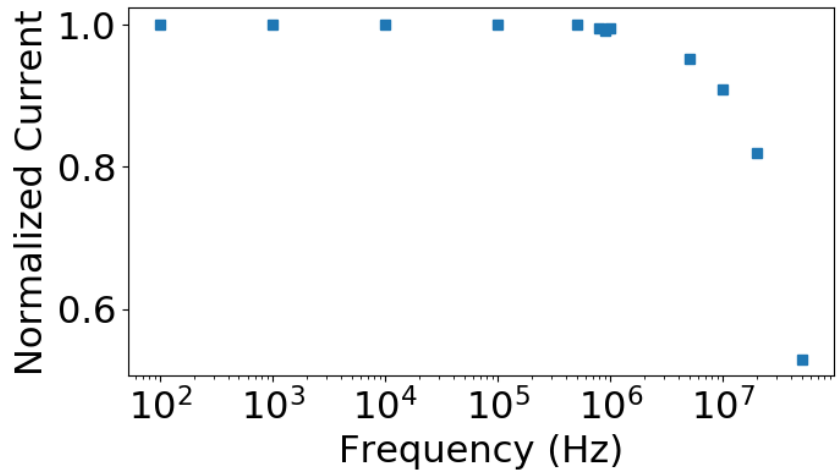

**Figure S7 | Frequency response of beam blanker.** Measured DC electron beam current through a Faraday cup with square wave blanking as a function of frequency using 2 mm blanking plates. The roll-off in current at high frequency is due to the RC time constant of the electrostatic blanking components. The current is normalized to DC current at 100 Hz blanking frequency. The theoretical DC current of the periodically blanked beam should be ½ that of the unblanked beam, which is the case for low frequencies. At 5 MHz, the current has dropped to approximately 95% of the current at 100 Hz.

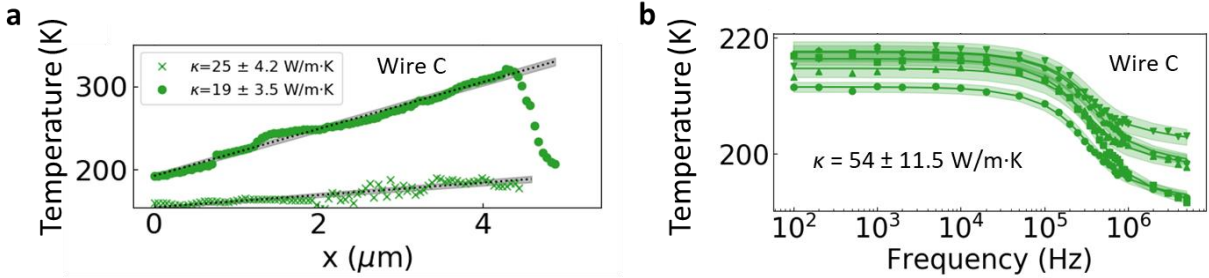

**Figure S8 | Cathodoluminescence thermal conductivity measurements on a nanowire with large doping variation. a**, Measurements using the DC slope method on Wire C, a nanowire with drastic variation in doping along its length, shown in Supplementary Fig. 6. "o" data points have an incident electron current of 9.3 nA, and "x" data points have an incident electron current of 3.2 nA. See discussions in the main paper related to aperturing the beam with regards to this data. **b**, AC method measurements to determine thermal conductivity on the same wire as in **a**. Unlike Wires A and B, Wire C has significant non-uniform doping along the wire (see Supplementary Fig. 6). We find that extracted thermal conductivities show larger discrepancies between the two methods. We attribute this to non-uniform thermal conductivity due to the doping variations, which invalidate the mathematical models used.

**Supplementary Note | DC bridge method derivation.**

We derive Equation 2 in the paper for the DC bridge method in the following manner. We use a thermal circuit model where $R = \frac{l}{\kappa A}$, where $R$ is thermal resistance, $l$ is length of segment, $\kappa$ is thermal conductivity of segment, and $A$ is cross-sectional area of segment. In a thermal circuit model, voltage is analogous to temperature difference, $\Delta T$, resistance is analogous to thermal resistance, and current is analogous to heat flux, $\dot{Q}$. The thermal circuit model equivalent of Ohm's law is then $\Delta T = \dot{Q}R$. We fix the temperature at the ends of the wire ($x = 0$ and $x = L$ where $L$ is the total wire length) as $T_0$. Because Pt partially coats the ends of the wire, as an approximation, we split the wire into 3 different regions. From $x = 0$ to $x = L_1$, the thermal conductivity is $\kappa_0$, a mix of the thermal conductivity of Pt and GaN (any effect due to increase of cross-sectional area of this region is incorporated into $\kappa_0$). From $x = L_1$ to $x = L_2$, the thermal conductivity is $\kappa_{GaN}$. From $x = L_2$ to $x = L$, the thermal conductivity is again $\kappa_0$. We neglect thermal contact resistance at the ends of the wires. If we apply a heat flux at location $x$, we can determine the temperature rise at $x$ by solving the thermal circuit model. We have to consider 3 different cases: when $0 \leq x \leq L_1$, when $L_1 \leq x \leq L_2$, and when $L_2 \leq x \leq L$, as we will need to solve parallel resistance equations and $l$ for each $R$ can change depending on where $x$ is, as will become clear.

For the case of $0 \leq x \leq L_1$, relevant parameters have been shown in Supplementary Fig. S9. To solve for the temperature rise at $x$, where we also have a heat flux $\dot{Q}$ from the incident electron beam, we need to solve for the total thermal resistance at point $x$. This appears as two parallel resistance paths to ground. One path is $R_A$, the other path is $R_B + R_C + R_D$, giving a total resistance of $R_{Total}(x) = \left( \frac{1}{R_A} + \frac{1}{R_B + R_C + R_D} \right)^{-1}$. We use $T(x) - T_0 = \Delta T(x) = \dot{Q}R_{Total}(x)$ to find the first line in Equation 2 in the paper. The other expressions for the cases of $L_1 \leq x \leq L_2$ and $L_2 \leq x \leq L$ can be found in a similar manner.

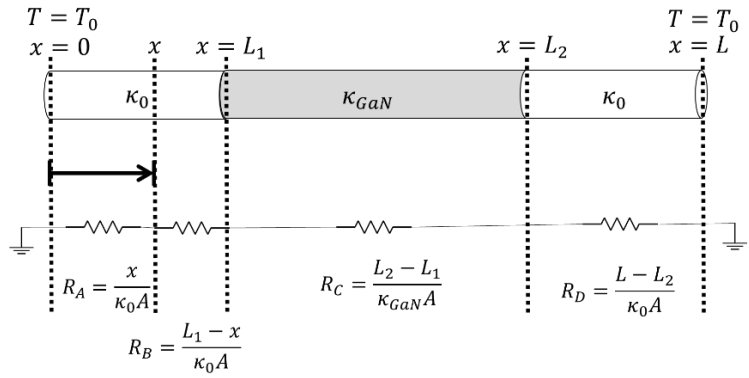

**Figure S9 | DC bridge method variables for $0 \leq x \leq L_1$.** Relevant parameters to compute the temperature profile in the DC bridge method are shown with description in the text. Heat flux $\dot{Q}$ is injected at $x$ from the electron beam.

When we fit our experimental data with Equation 2 in the manuscript, the fit parameters are $\kappa_{GaN}$, $\kappa_0$, $L_1$, and $L_2$.

**Supplementary Note | Time-dependent 1D temperature profile.**

To solve the time-averaged temperature at one end of the GaN wire during the "on" cycle of the electron beam as a function of frequency, we treat the wire as a 1D, uniform system to solve the dimensionless heat equation,

$$u_t = u_{xx}, \quad (S1)$$

where $u$ is temperature deviation from the overall average system temperature. Our dimensionless parameters are $\tilde{t} = \frac{t\kappa}{L^2 C_p \rho}$, $\tilde{x} = \frac{x}{L}$, and $\tilde{\omega} = \frac{\omega L^2 C_p \rho}{\kappa}$, where $L$ is wire length, $C_p$ is heat capacity, $\rho$ is

density, and $\kappa$ is thermal conductivity. Here the dimensionless parameter is used in the equations and

the tilde is dropped. The boundary conditions and initial condition are

$u(x = 0, t) = 0,$

$u_x(x = 1, t) = \eta \sum_{m=1,3,5,\dots}^{\infty} \frac{1}{m} \sin(m\omega t) = \eta \sum_{m=1,3,5,\dots}^{\infty} \frac{1}{m} (e^{im\omega t} - e^{-im\omega t}),$ \hfill (S2)

$u(x, t = 0) = 0.$

The time-dependent Neumann boundary condition is the Fourier series for a square wave heat input.

We expect the heat input to approximate a square wave for frequencies up to approximately 5 MHz

before the RC time constant of the electrostatic beam blanker alters the square wave shape (see Figure

S7). We expect a solution of the form

$$u(x, t) = v(x, t) + \sum_{m=1,3,5,\dots}^{\infty} B_m(x) \cos(m\omega t + \phi_m(x)).$$

Here, $\phi_m(x)$ is the phase term, $v(x,t)$ is the transient component and the second term is the quasi-steady state. We expect the system to reach the quasi-steady state on a much shorter time scale than the exposure time of the spectrometer (10's of ms or longer) due to the small size and heat capacity of the structure, so the average temperature we read on the spectrometer will be composed of only the second term. We convert this second term, $u_{qss}$, to the imaginary domain

$$u_{qss}(x,t) = Re\left\{\sum_{m=1,3,5,\dots}^{\infty} B_m(x)e^{i\phi_m(x)}e^{im\omega t}\right\} = Re\left\{\sum_{m=1,3,5,\dots}^{\infty} C_m(x)e^{im\omega t}\right\},$$

where $C_m$ is a grouping of all $x$-dependent terms. Using the identity $Re(z) = \frac{1}{2}(z + z^*)$, where $z$ is a complex number and $z^*$ is its complex conjugate, we find

$$u_{qss}(x,t) = \frac{1}{2}\left(\sum_{m=1,3,5,\dots}^{\infty} C_m(x)e^{im\omega t} + C_m^*(x)e^{-im\omega t}\right) = \sum_{m=1,3,5,\dots}^{\infty} B_m(x)\cos\left(m\omega t + \phi_m(x)\right) \qquad (S3)$$

Putting Equation S3 (center expression) into Equation S1, multiplying by 2, grouping terms, and dropping the sums due to the orthogonality of sines, we find

$$\left(im\omega C_m(x) - C_m''(x)\right)e^{im\omega t} - \left(im\omega C_m^*(x) + C_m''^*(x)\right)e^{-im\omega t} = 0.$$

Using the identity that if $ae^{i\omega t} + be^{-i\omega t} = 0$, then $a = b = 0$, we find

$$im\omega C_m(x) - C_m''(x) = 0 = -\left(im\omega C_m^*(x) + C_m''^*(x)\right). \qquad (S4)$$

We just need to solve one side of this equation, as the left is the complex conjugate of the right side. We then use the boundary conditions in Equation S2 applied to Equation S3 and find that $C_m(0) = 0$ and $C_m'(1) = \frac{A}{im}$. These are the boundary conditions needed to solve Equation S4. We find, noting that

$$i\omega m = \left(\sqrt{\frac{\omega m}{2}}(1+i)\right)^2,$$

$$C_m(x) = d_1 e^{-\sqrt{\frac{\omega m}{2}}(1+i)x} + d_2 e^{-\sqrt{\frac{\omega m}{2}}(1+i)x}$$

where $d_1$ and $d_2$ are unknowns we solve for with the $C_m(x)$ boundary conditions we just found. We determine

$$d_1 = -\frac{\frac{\eta}{im}}{\sqrt{\frac{\omega m}{2}}(1+i)e^{\sqrt{\frac{\omega m}{2}}(1+i)} + \sqrt{\frac{\omega m}{2}}(1+i)e^{-\sqrt{\frac{\omega m}{2}}(1+i)}}$$

$$d_2 = \frac{\frac{\eta}{im}}{\sqrt{\frac{\omega m}{2}}(1+i)e^{\sqrt{\frac{\omega m}{2}}(1+i)} + \sqrt{\frac{\omega m}{2}}(1+i)e^{-\sqrt{\frac{\omega m}{2}}(1+i)}},$$

thus,

$$C_m(x) = \frac{\eta}{im}\left(\frac{e^{\sqrt{\frac{\omega m}{2}}(1+i)x} - e^{-\sqrt{\frac{\omega m}{2}}(1+i)x}}{\sqrt{\frac{\omega m}{2}}(1+i)e^{\sqrt{\frac{\omega m}{2}}(1+i)} + \sqrt{\frac{\omega m}{2}}(1+i)e^{-\sqrt{\frac{\omega m}{2}}(1+i)}}\right),$$

and finally,

$$u_{qss}(x,t) = Re\left\{\sum_{m=1,3,5,\ldots}^{\infty}\frac{\eta}{im}\left(\frac{e^{\sqrt{\frac{\omega m}{2}}(1+i)x} - e^{-\sqrt{\frac{\omega m}{2}}(1+i)x}}{\sqrt{\frac{\omega m}{2}}(1+i)e^{\sqrt{\frac{\omega m}{2}}(1+i)} + \sqrt{\frac{\omega m}{2}}(1+i)e^{-\sqrt{\frac{\omega m}{2}}(1+i)}}\right)e^{im\omega t}\right\}. \quad (S5)$$

We only measure the temperature when the electron beam is on, or just half a period. Thus, we can integrate Equation S5 over a half period to find the average temperature,

$$\overline{u_{qss}\left(x, t=0..\frac{\pi}{\omega}\right)}$$

$$= \frac{\omega}{\pi}\int_0^{\pi/\omega} Re\left\{\sum_{m=1,3,5,\dots}^{\infty}\frac{\eta}{im}\left(\frac{e^{\sqrt{\frac{\omega m}{2}}(1+i)x}-e^{-\sqrt{\frac{\omega m}{2}}(1+i)x}}{\sqrt{\frac{\omega m}{2}}(1+i)e^{\sqrt{\frac{\omega m}{2}}(1+i)}+\sqrt{\frac{\omega m}{2}}(1+i)e^{-\sqrt{\frac{\omega m}{2}}(1+i)}}\right)e^{im\omega t}\right\}dt.$$

Solving, we find

$$\overline{u_{qss}\left(x, t=0..\frac{\pi}{\omega}\right)} = \frac{2\eta}{\pi}Re\left\{\sum_{m=1,3,5,\dots}^{\infty}\frac{1}{m^2}\left(\frac{e^{\sqrt{\frac{\omega m}{2}}(1+i)x}-e^{-\sqrt{\frac{\omega m}{2}}(1+i)x}}{\sqrt{\frac{\omega m}{2}}(1+i)e^{\sqrt{\frac{\omega m}{2}}(1+i)}+\sqrt{\frac{\omega m}{2}}(1+i)e^{-\sqrt{\frac{\omega m}{2}}(1+i)}}\right)\right\}.$$

Putting the dimensions back into the equation and solving at the end of the wire (x=L), with $\eta = \frac{4\dot{Q}x}{A\kappa\pi^2}$,

where $\dot{Q}$ is the power deposited by the electron beam (in Watts) and $A$ is wire cross-sectional area,

$$\overline{T_{qss}\left(L, t=0..\frac{\pi}{\omega}\right)}$$

$$= \frac{8\dot{Q}x}{A\kappa\pi^2}Re\left\{\sum_{m=1,3,5,\dots}^{\infty}\frac{1}{m^2}\left(\frac{e^{\sqrt{\frac{\omega L^2 C_p\rho m}{2\kappa}}(1+i)}-e^{-\sqrt{\frac{\omega L^2 C_p\rho m}{2\kappa}}(1+i)}}{\sqrt{\frac{\omega L^2 C_p\rho m}{2\kappa}}(1+i)e^{\sqrt{\frac{\omega L^2 C_p\rho m}{2\kappa}}(1+i)}+\sqrt{\frac{\omega L^2 C_p\rho m}{2\kappa}}(1+i)e^{-\sqrt{\frac{\omega L^2 C_p\rho m}{2\kappa}}(1+i)}}\right)\right\}$$

$$= \frac{8\dot{Q}x}{A\kappa\pi^2}Re\left\{\sum_{m=1,3,5,\dots}^{\infty}\frac{1}{m^2}\frac{\tanh\left(\sqrt{\frac{\omega L^2 C_p\rho m}{2\kappa}}(1+i)\right)}{\sqrt{\frac{\omega L^2 C_p\rho m}{2\kappa}}(1+i)}\right\}.$$

The full expression for the temperature we should measure with a square wave electron beam parked at the end of the wire, for temperature at the fixed end of $T_0$, is

$$\overline{T}_{meas}(\omega) = T_0 + \frac{4\dot{Q}L}{A\kappa\pi} + \overline{T_{qss}\left(L, t=0..\frac{\pi}{\omega}\right)}.$$